\documentclass[aps,letterpaper,nofootinbib,preprint,amsmath,amssymb,12pt]{revtex4}%
\usepackage{latexsym}%
\usepackage{epsf}%
\usepackage[hypertex]{hyperref}%

\def\pA{{\mathpzc A}}
\def\cF{{\mathcal F}}
\def\cH{{\mathcal H}}

\def\cN{{\mathcal N}}

\def\cW{{\mathcal W}}

\DeclareMathAlphabet{\mathpzc}{OT1}{pzc}{m}{it}

\newcommand{\beq}{\begin{equation}}
\newcommand{\beqn}{\begin{equation}\nonumber}
\newcommand{\eeq}{\end{equation}}
\newcommand{\bea}{\begin{eqnarray}}
\newcommand{\bean}{\begin{eqnarray}\nonumber}
\newcommand{\eea}{\end{eqnarray}}

\begin{document}

\begin{center}
{\bf{\Large Signatures of an Emergent Gravity}}
\vskip 3mm
{\bf{\Large from Black Hole Entropy}}
\bigskip
\bigskip

Cenalo Vaz
\bigskip

{\it RWC and Department of Physics,}\\
{\it University of Cincinnati,}\\
{\it Cincinnati, Ohio 45221-0011, USA}\\
{email: Cenalo.Vaz@UC.Edu}
\medskip

\end{center}
\bigskip
\medskip

\centerline{ABSTRACT}
\bigskip\bigskip
The existence of a thermodynamic description of horizons indicates that spacetime 
has a microstructure. While the ``fundamental'' degrees of freedom remain elusive, 
quantizing Einstein's gravity provides some clues about their properties. A quantum 
AdS black hole possesses an equispaced mass spectrum, independent of Newton's 
constant, $G$, when its horizon radius is large compared to the AdS length. Moreover, 
the black hole's thermodynamics in this limit is inextricably connected with its 
thermodynamics in the opposite (Schwarzschild) limit by a duality of the Bose 
partition function. $G$, absent in the mass spectrum, reemerges in the thermodynamic 
description through the Schwarzschild limit, which should be viewed as a natural 
``ground state''. It seems that the Hawking-Page phase transition separates fundamental, 
``particle-like'' degrees of freedom from effective, ``geometric'' ones.
\vskip 2.0in
	\vfill\eject

The degrees of freedom responsible for the dramatic gain in the entropy of a matter cloud when 
it collapses to form a black hole \cite{bek72,bch73,haw75}, are generally attributed 
to quantum gravity. However, it has become clear that significantly different approaches to 
quantum gravity are able to successfully reproduce the black hole entropy from a microcanonical 
or a canonical ensemble of ``elementary'' degrees of freedom. This multiplicity of successes 
obscures the true nature of the ``atom'' of spacetime. 

There are, nevertheless, significant variations in the details of the calculations carried out in the 
various approaches to quantum gravity, a most curious one being in the use of statistics. 
The microstates of string theory \cite{stva96,hrpl97,dab05} and AdS/CFT
\cite{hkl99,emp99,ahaoo00,hmst01,mupa02,chgs07} are assumed indistinguishable whereas those of Loop 
Quantum Gravity (LQG) \cite{rov96,abck97,dle04,mei04,cor07,bar08} must be distinguishable to reproduce 
the area law. The issue of counting has been difficult to address because the relationships between 
approaches are poorly understood. Moreover, LQG is yet to deal with AdS black holes and the field 
theoretic approaches are uncomfortable with the Schwarzschild black hole. What is needed is an approach 
that, although possibly not as fundamental, is able to address both.

Our recent canonical quantization of spherical dust collapse in AdS space has provided 
important insights into questions concerning Hawking radiation, the information loss paradox, the 
nature of the black hole eigenspectrum and black hole entropy \cite{va1,va2,va3,va4}.
In this essay we will discuss the role of statistics in state counting and speculate on 
its implications. 

The classical spherical collapse of inhomogeneous dust in dimension $d=n+2$ is described by the 
LeMaitre-Tolman-Bondi (LTB) family of metrics. The models may be expressed in canonical form 
after a series of simplifying canonical transformations and after absorbing the surface terms 
\cite{kuc94,kas94,va5,va6}. One finds that they are described in the phase space consisting of 
the dust proper time, $\tau(t,r)$, the area radius, $R(t,r)$, the mass density, $\Gamma(r)$, and 
their conjugate momenta, $P_\tau(t,r)$, $P_R(t,r)$ and $P_\Gamma(t,r)$ respectively, by the two 
constraints \cite{va3}
\bea
&&\cH_r = \tau'P_\tau + R' P_R -\Gamma P_\Gamma' \approx 0\cr\cr
&&\cH^{g}=P_\tau^2 + \cF P_R^2 -\frac{\Gamma^2}{\cF}\approx 0,
\label{const1}
\eea
where $t$ and $r$ are ADM label coordinates, the primes refer to derivatives with respect to $r$, 
$\Gamma(r)$ is defined as the space derivative of the mass function, $F(r)$, of the collapsing dust 
and $\cF$ is given by
\beq
\cF = 1-\frac F{R^{n-1}}+\frac{2R^2}{n(n+1)l^2},
\label{cF}
\eeq
with $\Lambda = -l^{-2}$ representing the cosmological constant. Dirac quantization leads 
to a simplified Wheeler-DeWitt equation which, for a smooth dust distribution, can be regularized 
on a lattice. Assuming that the wave-functional is factorizable, it can quite generally be taken 
of the form
\beq
\Psi[\tau,R,\Gamma] = \exp\left[i\int dr \Gamma(r) \cW(\tau(r),R(r),F(r))
\right]
\label{wfnal}
\eeq
and automatically obeys the momentum constraint provided that $\cW(\tau,R,F)$ has no explicit 
$r-$dependence. It turns out that the wave-functional must satisfy three equations \cite{va6}, 
one of which is the Hamilton--Jacobi equation and was used to describe Hawking radiation in 
\cite{va3,va7}. The other two equations together uniquely fix the Hilbert space measure and 
the factor ordering.

Eternal black holes, which are the central concern of this essay, are special solutions within the 
LTB family, obtained when the mass function is constant and the energy function is vanishing.
Thus the mass density function is distributional and the wave functional in \eqref{wfnal} 
turns into a wave-{\it function}. In contrast to the situation in which the mass density is
smooth, no regularization is required but there is an ambiguity in both the factor ordering and 
Hilbert space measure. However, we showed that a successful description of the Hawking evaporation 
of a collapsing dust cloud surrounding a pre-existing black hole depends crucially on the correct
choice of measure for eternal black holes \cite{va3}. This measure, which we adopt here, is 
the one obtained from the DeWitt supermetric and also uniquely fixes the appropriate factor ordering 
\cite{va4}. Eternal black holes become described by the free Klein-Gordon equation
\beq
\left[\frac{\partial^2}{\partial \tau^2} \pm \frac{\partial^2}{\partial R_*^2}\right] \Psi = 0,
\eeq
where the positive sign refers to the exterior, the negative sign to the interior and $R_*$ 
is
\beq
R_*(R,M,l) = \pm \int \frac{dR}{\sqrt{|\cF|}}.
\eeq
Because the wave equation is hyperbolic only in the interior, the black hole wave function is 
supported only there. Standard boundary conditions at the horizon then yield a spectrum of the 
form
\beq
\pA=\frac{n\Omega_n}{4\pi}~ (2G_d M) L_h = A_\text{Pl} \left(j+\frac 12\right),
\label{quant1}
\eeq
where $j$ is a whole number, $A_\text{Pl}=h G_d$ is the Planck area and $L_h$ is the proper 
radius of the horizon. It is not possible to give an analytical expression for $L_h$ in general.

$\pA$ becomes the horizon area when the dimensionless variable $x_h = R_h/l$, where $R_h$ 
is the horizon radius, satisfies $2x_h^2 \ll n(n+1)$. In this limit one then finds an equispaced 
area spectrum
\beq
A_j = \frac{4G_dh\sqrt{\pi}\Gamma\left(\frac 1{n-1}\right)}{n\Gamma\left(\frac 1{n-1}+\frac 12\right)}
\left( j + \frac 12\right),
\label{areaquant}
\eeq
where $j$ is a whole number. We will refer to it as the Schwarzschild limit. The spectrum 
in \eqref{areaquant} is compatible with results from LQG for large mass black holes. On the other 
hand, $L_h$ is independent of the mass in the opposite limit, $2x_h^2 \gg n(n+1)$ and \eqref{quant1} 
predicts an equispaced {\it mass} spectrum
\beq
M_j = \frac{4h}{nl\Omega_n}\sqrt{\frac{n+1}{2n}} \left(j+\frac 12\right),
\label{massquant}
\eeq
with levels that are independent of the gravitational constant.

We will now discuss black hole thermodynamics in each limit, but first it is necessary to specify 
what we mean by a black hole microstate.
From what we have said earlier, the black hole can be viewed as a single shell with the spectrum 
in \eqref{quant1}. However, this single shell is in fact the end state of many shells that have 
collapsed to form the black hole. If we assume that, regardless of their history, each of the shells 
then occupies only the levels of \eqref{quant1}, a black hole microstate becomes a particular 
distribution of collapsed shells among the available levels and the black hole itself should 
properly be interpreted as an excitation by $\cN=\sum_j \cN_j$ collapsed shells. 

With the area spectrum in \eqref{areaquant}, one  can recover the Bekenstein-Hawking entropy by a 
direct counting of states in an ``area ensemble'' provided that they are assumed distinguishable 
(Boltzmann statistics) \cite{va4}. The result is,
\beq
S = f(n) \frac A{4 A_\text{Pl}},
\label{entmicro}
\eeq
where $f(n)$ is dimension dependent and approximately equal to one. The fact that the ``area'' 
quanta must be treated as distinguishable runs contrary to our intuition for elementary 
degrees of freedom in quantum field theory and calls into question whether ``area'' is fundamental 
in quantum gravity as suggested in \cite{pad03}. 
 
In the opposite limit and with the mass spectrum in \eqref{massquant}, Bose statistics must be 
employed and the partition function describing the black hole is 
\beq
Z(\xi) = \prod_{j=0}^\infty\left[1-e^{-\xi(2j+1)}\right]^{-1},~~ \xi=\frac{2h \beta}{nl\Omega_n}
\sqrt{\frac{n+1}{2n}}.
\label{partfun}
\eeq
A well-known duality \cite{hr18}, relates $Z(\xi)$ and $Z(\xi^{-1})$ according to \cite{va1}, 
\beq
Z(\xi) = \prod_{j=0}^\infty \left[1-e^{-\xi(2j+1)}\right] = \frac 1{\sqrt{2}}e^{\frac{\pi^2}{12\xi}
+\frac{\xi}{24}}[Z(2\pi^2/\xi)]^{-1}.
\label{duality}
\eeq
If we assume that $\xi \ll 1$ (large surface gravity), the value of $Z(2\pi^2/\xi)$ will depend 
on assumptions concerning the ``ground state'' of the system. We do not {\it \`a priori} know the 
partition function of this ground state, but we do know that whatever our choice of $Z(2\pi^2/\xi)$, 
the partition function in \eqref{partfun} must yield the Bekenstein-Hawking entropy. This follows 
because we are quantizing Einstein's gravity, whose equivalence with the Bekenstein-Hawking area law 
has been established \cite{jac95,pad07}.  Strikingly, it turns out \cite{va4} that we must choose,
for $Z(2\pi^2/\xi)$, the partition function in the Schwarzschild limit as obtained from \eqref{entmicro}. 
This identifies the Schwarzschild limit as the ``ground state'', through which $G$ enters into 
the thermodynamics of ``large'' black holes. 

The two limits considered here are separated by the Hawking-Page phase transition \cite{hawpag83}, 
so they are two different phases of the same thermodynamic system. The large horizon limit is more 
in keeping with quantum field theory both because the heat capacity is positive and because the 
fundamental ``particles'' must be assumed indistinguishable. The phase transition therefore 
describes a change in the degrees of freedom from ``field theoretic'' ones to ``geometric'' ones, 
which must be assumed distinguishable and for which the heat capacity is negative.


\begin{thebibliography}{99}
\bibitem{bek72}J. D. Bekenstein, Ph.D. thesis, Princeton University (1972); {\it ibid} Lett. 
Nuovo Cimento {\bf 4} (1972) 737; {\it ibid} Phys. Rev. D {\bf 7} (1973) 2333.
\bibitem{bch73}J. M. Bardeen, B. Carter, and S. W. Hawking,
  Comm. Math. Phys. {\bf 31} (1973) 161.
\bibitem{haw75}S. W. Hawking, Comm. Math. Phys. {\bf 43} (1975) 199.
\bibitem{stva96}A. Strominger and C. Vafa, Phys. Lett. B {\bf 379} (1996) 99.
\bibitem{hrpl97}G.T. Horowitz and J. Polchinski, Phys. Rev. D {\bf 55}
  (1997) 6189. 
\bibitem{dab05}A. Dabholkar, Phys. Rev. Lett. {\bf 94} (2005) 241301.
\bibitem{hkl99}S. Hyun, W.T. Kim, and J. Lee, Phys. Rev. D {\bf 59}
  (1999) 084020. 
\bibitem{emp99}R. Emparan, JHEP 9906 (1999) 036.
\bibitem{ahaoo00} O. Aharony, S. S. Gubser, J. M. Maldacena, H. Ooguri, and 
Y. Oz, Phys. Rept. {\bf 323} (2000) 183.
\bibitem{hmst01}S. W. Hawking, J. Maldacena, and A. Strominger, JHEP 0105
  (2001) 001. 
\bibitem{mupa02}S. A. Mukherji and S. S. Pal, JHEP 0205 (2002) 026.
\bibitem{chgs07}S. K. Chakrabarti, K.S. Gupta, and S. Sen,
 Int. J. Mod. Phys. A {\bf 23} (2008) 2547.
\bibitem{rov96}C. Rovelli, Phys. Rev. Lett. {\bf 77} (1996) 3288.
\bibitem{abck97}A. Ashtekar, J. Baez, A. Corichi, and K. Krasnov,
  Phys. Rev. Lett. {\bf 80} (1998) 904. 
\bibitem{dle04}M. Domagala and J. Lewandowski, Class. Quant. Grav. {\bf 21} (2004) 5233. 
\bibitem{mei04}K. A. Meissner, Class. Quant. Grav. {\bf 21} (2004) 5245.
\bibitem{cor07} A. Corichi, J. Di\'az-Polo and E. Fern\'andez-Borja, Phys. Rev. Lett. 
{\bf 98} (2007) 181301.
\bibitem{bar08} I. Agull\'o, J. F. Barbero G., J. Di\'az Polo, E. Fern\'andez-Borja 
and E. J. S. Villase\v nor, Phys. Rev. Lett. {\bf 100} (2008) 211301.
\bibitem{va1} C. Vaz, S. Gutti, C. Kiefer, T.P. Singh and L.C.R. Wijewardhana, 
Phys. Rev. D {\bf 77} (2008) 064021. 
\bibitem{va2} R. Tibrewala, S. Gutti, T.P. Singh and C. Vaz, Phys. Rev D {\bf 77} 
(2008) 064012.
\bibitem{va3} C. Vaz, R. Tibrewala and T.P. Singh, Phys. Rev. D {\bf 78} (2008) 024019. 
\bibitem{va4} C. Vaz and L.C.R. Wijewardhana, Phys. Rev. D {\bf 79} (2009) 084014.
\bibitem{kuc94} K. Kucha\v r, Phys. Rev. D {\bf 50} (1994) 3961.
\bibitem{kas94} H. Kastrup and T. Thiemann, Nucl. Phys. B {\bf 425} (1994) 665.
\bibitem{va5} C. Vaz, L. Witten and T.P. Singh, Phys. Rev. D {\bf 63} (2001) 104020.
\bibitem{va6} C. Kiefer, J. Mueller-Hill, C. Vaz, Phys. Rev. D {\bf 73} (2006) 044025.
\bibitem{pad03} T. Padmanabhan, Phys. Rept. {\bf 406} (2005) 49-125.
\bibitem{va7} C. Vaz, C. Kiefer, T.P. Singh and L. Witten, Phys. Rev. D {\bf 67} 
(2003) 0204014. 
\bibitem{hr18}G. Hardy and S. Ramanujan, Proc. Lond. Math. Soc. {\bf 17} (1918) 75.
\bibitem{jac95} T. Jacobson, Phys. Rev. Lett. {\bf 75} (1995) 1260.
\bibitem{pad07} T. Padmanabhan, AIPConf. Proc. 939 (2007) 114, 
\href{http://arxiv.org/abs/0706.1654}{[arXiv:0706.1654]}.
\bibitem{hawpag83} S. W. Hawking and D. N. Page, Commun. Math. Phys. {\bf 87} (1983) 
577.
\end{thebibliography}
\end{document}